\documentclass[conference]{IEEEtran}
\IEEEoverridecommandlockouts
\usepackage{balance}
\usepackage{cite}
\usepackage{amsmath,amssymb,amsfonts}
\usepackage{array}
\usepackage{stfloats}
\usepackage[caption=false,font=footnotesize,position=bottom]{subfig}
\usepackage{url}
\usepackage{graphicx}
\usepackage{textcomp}
\usepackage{xcolor}
\usepackage{tikz}
\usepackage{tikz-3dplot}
\usepackage{ellipsis}
\usetikzlibrary{calc}
\usetikzlibrary{decorations.pathreplacing,decorations.markings,shapes.geometric}
\usetikzlibrary{calc,patterns,angles,quotes,shapes,arrows.meta}
\usetikzlibrary{chains,spy}
\tikzset{>=Stealth}
\usepackage{cite}
\usepackage{amsmath,amssymb,amsfonts}
\usepackage[capitalise]{cleveref}
\usepackage{amsthm}

\usepackage{algorithm}
\usepackage{algpseudocode}
\usepackage{siunitx}

\usepackage[utf8]{inputenc}

\usepackage{mathtools}
\usepackage{bm}
\usepackage{etoolbox}
\usepackage{scalerel}

\usepackage{fancyhdr}

\newcommand{\vb}{{\bm{b}}}
\newcommand{\vc}{{\bm{c}}}

\newcommand{\vh}{{\bm{h}}}

\newcommand{\vn}{{\bm{n}}}

\newcommand{\vr}{{\bm{r}}}

\newcommand{\vv}{{\bm{v}}}

\newcommand{\vy}{{\bm{y}}}
\newcommand{\vz}{{\bm{z}}}

\newcommand{\mc}{{\bm{C}}}

\newcommand{\mq}{{\bm{Q}}}

\newcommand{\vth}{{\bm{\theta}}}
\newcommand{\vsig}{\bm{\sigma}}

\newcommand{\vmu}{\bm{\mu}}

\newcommand{\vphi}{{\bm{\phi}}}
\newcommand{\vpsi}{{\bm{\psi}}}

\newcommand{\diag}{\mathrm{diag}}

\newcommand{\He}{\mathrm{H}}
\newcommand{\T}{\mathrm{T}}

\newcommand{\I}{\mathbf{I}}

\newcommand{\NC}{\mathcal{N}_\mathbb{C}}

\newcommand{\E}{\mathbb{E}}
\renewcommand{\mid}{\, | \,}

\usepackage{amsthm}

\tikzstyle{block} = [draw, rectangle, 
minimum height=4em, minimum width=4em]
\tikzstyle{input} = [coordinate]
\tikzstyle{output} = [coordinate]
\tikzstyle{pinstyle} = [pin edge={to-,thin,black}]

\usetikzlibrary{positioning}

\tikzset{radiation/.style={{decorate,decoration={expanding waves,angle=90,segment length=5pt}}}}

\usetikzlibrary{spy}
\usepackage{pgfplots}
\usepackage{wrapfig}
\usetikzlibrary{arrows,shapes}
\usetikzlibrary{positioning,shapes.callouts}
\usepgfplotslibrary{groupplots,dateplot}
\usetikzlibrary{patterns,shapes.arrows}
\pgfplotsset{compat=newest}

\def\BibTeX{{\rm B\kern-.05em{\sc i\kern-.025em b}\kern-.08em
		T\kern-.1667em\lower.7ex\hbox{E}\kern-.125emX}}

\usepackage[acronym,shortcuts]{glossaries}
\newacronym{GMM}{GMM}{Gaussian mixture model}
\newacronym{CGMM}{CGMM}{coupled Gaussian mixture model}
\newacronym{cVAE}{cVAE}{conditional variational autoencoder}
\newacronym{VAE}{VAE}{variational autoencoder}
\newacronym{PDF}{PDF}{probability density function}
\newacronym{MSE}{MSE}{mean square error}
\newacronym{CSI}{CSI}{channel state information}
\newacronym{CME}{CME}{conditional mean estimator}
\newacronym{ML}{ML}{maximum likelihood}
\newacronym{LS}{LS}{least squares}
\newacronym{LOS}{LoS}{line-of-sight}
\newacronym{NLOS}{NLoS}{non-\ac{LOS}}
\newacronym{DoA}{DoA}{direction-of-arrival}
\newacronym{DoD}{DoD}{direction-of-departure}
\newacronym{SNR}{SNR}{signal-to-noise ratio}
\newacronym{TNR}{TNR}{time-to-noise ratio}
\newacronym{BS}{BS}{base station}
\newacronym{JCAS}{JCAS}{joint communication and sensing}
\newacronym{MIMO}{MIMO}{multiple-input-multiple-output}
\newacronym{MISO}{MISO}{multiple-input-single-output}
\newacronym{MMSE}{MMSE}{minimum \ac{MSE}}
\newacronym{NMSE}{NMSE}{normalized \ac{MSE}}
\newacronym{RMSE}{RMSE}{root \ac{MSE}}
\newacronym{LMMSE}{LMMSE}{linear minimum \ac{MSE}}
\newacronym{MT}{MT}{mobile terminal}
\newacronym{UE}{UE}{user equipment}
\newacronym{OMP}{OMP}{orthogonal matching pursuit}
\newacronym{CS}{CS}{compressed sensing}
\newacronym{ULA}{ULA}{uniform linear array}
\newacronym{DFT}{DFT}{Discrete Fourier Transform}
\newacronym{CRB}{CRB}{Cramer-Rao bound}
\newacronym{MUSIC}{MUSIC}{multiple signal classification}
\newacronym{rMUSIC}{rMUSIC}{root multiple signal classification}
\newacronym{ESPRIT}{ESPRIT}{estimation of signal parameters via rotational invariance techniques}
\newacronym{GE}{GE}{gridded estimator}
\newacronym{AWGN}{AWGN}{additive white Gaussian noise}
\newacronym{OFDM}{OFDM}{orthogonal frequency division multiplexing}
\newacronym{ADC}{ADC}{analog-to-digital converter}
\newacronym{wlog}{w.l.o.g.}{without loss of generality}
\newacronym{mmWave}{mmWave}{millimeter wave}
\newacronym{MAP}{MAP}{maximum a-posteriori}
\newacronym{PBCE}{PBCE}{parametric Bayesian channel estimator}
\newacronym{CGLM}{CGLM}{conditional Gaussian latent model}
\newacronym{AB}{AB}{asymptotic bound}
\newacronym{AP}{AP}{access point}
\newacronym{SPWMMSE}{SPWMMSE}{stochastic proximal weighted minimum mean square error}
\newacronym{WMMSE}{WMMSE}{weighted minimum mean square error}
\newacronym{SWMMSE}{SWMMSE}{stochastic weighted minimum mean square error}
\newacronym{UL}{UL}{uplink}
\newacronym{DL}{DL}{downlink}
\newacronym{MU}{MU}{multi-user}
\newacronym{FDD}{FDD}{frequency divison duplex}
\newacronym{SR}{SR}{sum rate}
\newacronym{cCDF}{cCDF}{complementary cumulative density function}
\newacronym{CNN}{CNN}{convolutional neural network}
\newacronym{NN}{NN}{neural network}
\newacronym{GPS}{GPS}{global positioning system}
\newacronym{GNSS}{GNSS}{global navigation satellite system}
\newacronym{ELBO}{ELBO}{evidence-lower bound}
\newacronym{KL}{KL}{Kullback-Leibler}
\newacronym{DNN}{DNN}{deep neural network}

\tikzset{PlotDFT/.style={mark=star,mark size=2.2pt, line width=1pt, color=gray, dashed, mark options=solid}}
\tikzset{PlotCVAE-single/.style={mark=triangle,mark size=1.5pt, line width=1pt, color=green!40!black, dashed, mark options={solid, rotate=180}}}
\tikzset{PlotGeo/.style={mark=triangle,mark size=1.5pt, line width=1pt, color=black, dashed, mark options={solid, rotate=0}}}
\tikzset{PlotCVAE-no-cond/.style={mark=diamond,mark size=1.8pt, line width=1pt, dashed, color=black!50!green, mark options=solid}}
\tikzset{PlotCVAE-genie/.style={mark=x, mark size=2.2pt, domain=1:10000, line width=1pt, color=red, dashed, mark options=solid}}
\tikzset{PlotCVAE-cond/.style={mark=square,mark size=1.5pt, line width=1pt, color=blue, mark options=solid}}
\tikzset{PlotCGMM-S/.style={mark=o, mark size=1.5pt, line width=1pt, color=orange , dashed, mark options=solid}}

\usepackage{circuitikz}

\usepackage{fancyhdr}

\fancypagestyle{cfooter}{ %
	\fancyhf{} 
	\cfoot{	\small This work has been submitted to the IEEE for possible publication. Copyright may be transferred without notice, after which this version may no longer be accessible.
	}
	
}

\begin{document}

\title{
Context-Aware CSI Prediction for Access Point Selection Utilizing Conditional VAEs
\thanks{This work is funded by the Bavarian Ministry of Economic Affairs, Regional Development, and Energy within the project 6G Future Lab Bavaria.}
}
\newcommand{\jbig}{$\mathcal{J}$}
\author{\IEEEauthorblockN{Franz Weißer, Amar Kasibovic, and Wolfgang Utschick\\}
\IEEEauthorblockA{\textit{TUM School of Computation, Information and Technology, Technical University of Munich, Germany} \\
\{franz.weisser, amar.kasibovic, utschick\}@tum.de}
}

\maketitle

\thispagestyle{cfooter}

\begin{abstract}
    Indoor wireless communication environments are strongly influenced by dynamic conditions, which affect channel state information (CSI) and, consequently, the precoding strategy and the selection of the access point (AP).
    Device-free sensing and localization functionalities can provide information about these conditions, including, for example, the user's position and the position of mobile blocking objects.
    To model the statistical relationship between the CSI and the provided conditions, we employ a conditional variational autoencoder (cVAE).
    We treat the user and object positions — referred to as context information — as conditional inputs to the cVAE.
    The proposed model does not rely on ground-truth CSI and is trained directly on noisy data.
    Once trained, the framework can infer channel statistics solely from user and blocking object positions, enabling proactive AP selection based on inferred statistical CSI without requiring continuous CSI estimation.
    Extensive simulations with the state-of-the-art ray-tracing tool Sionna validate the proposed method.
\end{abstract}


\section{Introduction}

Future systems will utilize a higher density of \acp{AP} to meet the needs of an increasing number of connected mobile \acp{UE} while remaining energy-efficient.
In such dynamic communication systems, knowledge of the \ac{CSI} is crucial for achieving a reliable connection and, consequently, high data rates.
More specifically, the multiple possible \acp{AP} need information about their communication link to the served \ac{UE} to design and update their transmit strategy.
The \ac{CSI} is highly dependent on current radio propagation characteristics and the \ac{UE}'s position, making the interplay between communication and sensing functionalities a core enabler for such dense systems.

Typically, based on the \ac{CSI} and other context information, the \ac{UE} is connected to one of the \acp{AP}, cf.~\cite{Hussain2025, Ohori2023}.
The \ac{UE}’s position may be provided by radar, lidar, or \ac{JCAS} functionalities, or, if the deployment is outdoor, by \ac{GNSS}.
The \ac{JCAS} functionalities also support device-free sensing of moving obstacles, which may significantly influence the current \ac{CSI} but are not part of the communication system.
In time-varying scenarios, i.e., when the \ac{UE} and/or obstacles are moving, the \ac{CSI} at the \acp{AP} must be updated regularly to ensure an optimal connection for the \ac{UE}.
Furthermore, 
anticipating the optimal serving \ac{AP} ahead of time
is crucial for ensuring continuous service to the \ac{UE} and meeting the quality of service demands of the communication system.
Hence, many works have been dedicated to predicting such \emph{handovers} based on a set of contextual information to prepare the \acp{AP} in advance and ensure a reliable connection, e.g.,~\cite{Hussain2025, Ohori2023}.
Such context information can include, but is not limited to, the \ac{UE}'s position, speed, and the position of any dynamic blocking obstacles.

Additionally, position-based radio maps have been developed, e.g., in~\cite{Levie2021, Zeng2021}.
Here, the path loss and, consequently, the \ac{SNR} for each \ac{UE} position and \ac{AP} are estimated, which enables \ac{SNR}-based \ac{AP} selection.
Furthermore, position-based precoding strategies for each respective \ac{AP} have been introduced, e.g.,~\cite{Maiberger2010,Miao2021,Weisser2025c}.
However, in dynamic scenarios characterized by continuously changing environments, these methods fail to adapt to the current radio propagation characteristics.
Hence, we propose a new method that adapts to such dynamic scenarios while providing statistical \ac{CSI} to perform localization- and sensing-aware precoding and \ac{AP} selection.

\emph{Contributions:}
Building on the recent idea of position-based precoding from~\cite{Weisser2025c}, which uses a multi-view generative model, we develop a framework based on the \ac{cVAE} that provides channel statistics solely from the \ac{UE}'s position and other context information.
To this end, a \ac{cVAE} is trained on noisy training samples to infer conditional channel statistics.
These learned channel statistics are not only used to select the optimal \ac{AP} but also to provide linear precoders for downlink transmission.
We empirically demonstrate the benefit of learning the conditional channel statistics from all available context information.
\section{Preliminaries}

\subsection{System Model}

We consider an indoor \ac{MISO} system with multiple \acp{AP}, where linear precoding is adopted in the \ac{DL}.
The $s$-th \ac{AP}, with $s\in\mathcal{S}=\{1,\dots,S\}$, is equipped with $N_s$ antennas.
The intended signal $x_s \in \mathbb{C}$ is linearly precoded with $\vv_s\in \mathbb{C}^{N_s}$.
The received signal at the \ac{UE} from the $s$-th \ac{AP} can be written as
\begin{align}
    y_s = \vh^{\He}_s\vv_s x_s + n, 
\end{align}
where $\vh_s$ and $n$ denote the $s$-th \ac{AP}-\ac{UE} channel and the \ac{AWGN}, respectively.
We assume that each \ac{AP} serves its \ac{UE} on a different carrier, meaning that we neglect inter-\ac{AP} interference.
Further, we assume $\E[x_s]=0$, $\E[x_s x_s^*]=1$, and $n\sim\NC(0,\sigma^2)$.
The resulting rate for the $s$-th \ac{AP} is then given as
\begin{align}
    R_s = \log_2\left(1 + \frac{\rho_s|\vh^\He_{s}\vv_{s}|^2}{\sigma^2 \|\vv_s\|^2 }\right), 
\end{align}
where $\rho_s$ is the maximum transmit power available at \ac{AP} $s$.
The overall goal is to maximize the \ac{UE}'s rate by selecting the \ac{AP} $s$ that yields the highest achievable rate $R_{s}$.
If multiple \acp{UE} are present, the \ac{AP} selection problem admits a game-theoretic formulation, specifically as a potential game~\cite{Mai2015}. 
Nevertheless, \ac{CSI} prediction for individual \acp{UE} remains relevant 
even in the multi-user case.

Full knowledge of the true underlying highly dimensional channel $\vh_s$ is infeasible in practical systems.
Hence, only noisy pilot observations of the wireless channel can be utilized for any downstream task.
We consider a system using $N_s$ downlink pilot symbols, resulting in noisy \ac{CSI} observations
\begin{align}
    \tilde{\vy}_s = \vh_s + \vn    
\end{align}
with $\vn\sim\NC(\bm{0},\sigma^2\I)$.

In this work, we not only assume imperfect \ac{CSI} knowledge but also infer channel statistics solely based on the respective \ac{UE} positions and other context information, jointly denoted by $\vc = [\vr^\T,\vb^\T]^\T$, where $\vr$ denotes the position vector of the \ac{UE} and $\vb$ denotes the position vector of a blocking object.
One should note that the method can be straightforwardly extended to include more context information in $\vc$.

 \begin{figure}
\centering
    \includegraphics[]{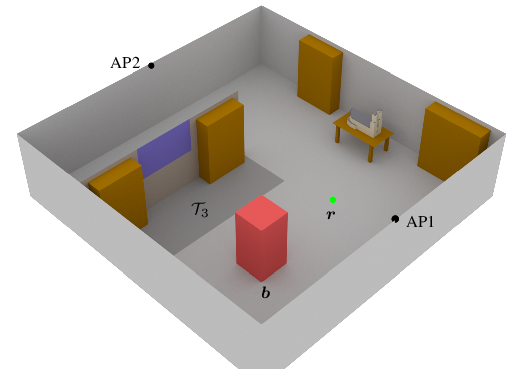}
\caption{Visualization of the indoor channel model with two \acp{AP} (black points), one possible \ac{UE} position (green point), and a possible location of the blocking object (red cuboid). The gray area corresponds to test set $\mathcal{T}_3$.}
    \label{fig:sionna_indoor}
\end{figure}

\subsection{Indoor Channel Model}
\label{sec:indoor}

We evaluate our approach by simulating an indoor room environment using the ray-tracing tool Sionna~\cite{Hoydis2022}. 
The room is $\qty{10}{m}$ wide, $\qty{10}{m}$ long, and $\qty{2.5}{m}$ high. 
Two \acp{AP} are mounted in the middle of two opposite walls at a height of $\qty{2.4}{m}$.
Furthermore, a moving object is simulated as a tall cuboid with a square cross-section of $\qty{1}{m^2}$ and a height of $\qty{2}{m}$.
We simulate at a carrier frequency of $\qty{28}{GHz}$.
The \acp{AP} are equipped with a \ac{ULA} facing towards the room center, i.e., the antenna array is positioned horizontally along the wall, consisting of $N_1 = N_2 = 32$ antennas with half-wavelength spacing.
\cref{fig:sionna_indoor} shows a visualization of the scene.
We assume that the \ac{AP} uses full power, and we define the \acp{UE}' receive \ac{SNR} corresponding to the $s$-th \ac{AP} as
\begin{align}
    \mathrm{SNR}_s = \frac{\rho_s\mathbb{E}[\|\vh_s\|^2]}{N_s\sigma^2}.
\end{align}

\section{Method}

In contrast to~\cite{Ohori2023, Hussain2025}, we do not assume any real-time \ac{CSI} knowledge in order to perform \ac{AP} selection but rather focus on learning a model that provides us with statistical CSI, which refers to a parameterized conditional \ac{PDF} $p^{(s)}(\vh_s\mid\vc)$ for each \ac{AP} $s\in\mathcal{S}$.
After building such a model, we can use it to predict the statistical \ac{CSI} corresponding to previously unseen contextual information.
We train the model on a dataset $\mathcal{T}_\T=\{(\{\tilde{\vy}_s^\ell\}_{s\in\mathcal{S}},\vc^\ell)\}_{\ell=1}^L$ comprising $L$ samples, where each sample consists of $\lvert \mathcal{S} \rvert$ noisy pilot observations -- one per \ac{AP} -- together with the corresponding context information $\vc^\ell$.

\subsection{Conditional Variational Autoencoder}

As we assume that the \acp{AP} have access to the \ac{UE}'s and obstacle's positions, i.e., the context information $\vc$, 
we 
use a generative model that provides conditional statistics $p^{(s)}(\vh_s\mid\vc;\vth_s)$,
where $\vth_s$ are the parameters of the model.
Such a model is the \ac{cVAE}~\cite{Sohn2015}, which is based on a continuous latent variable $\vz\in\mathbb{C}^{N_\mathrm{L}}$.
Here, the conditional log-likelihood of the noisy \ac{CSI} observations is lower bounded as
\begin{align}
    \log p^{(s)}(\tilde{\vy}_s \mid \vc;\vth_s) &\geq \mathbb{E}_{q^{(s)}(\vz\mid\tilde{\vy}_s,\vc;\vphi_s)}\left[\log p^{(s)}(\tilde{\vy}_s\mid \vz,\vc;\vth_s) \right] \nonumber\\
    &\hspace{-5pt}- \mathrm{D_{KL}} \big(q^{(s)}(\vz\mid\tilde{\vy}_s,\vc;\vphi_s) \,\big\|\, p^{(s)}(\vz\mid\vc;\vpsi_s)\big)\,,\label{eq:elbo_h}
\end{align}
where $\vth_s$, $\vphi_s$, and $\vpsi_s$ are the model parameters of the respective distributions.
The right-hand side of~\eqref{eq:elbo_h} is the so-called \ac{ELBO}, with the last term denoting the \ac{KL} divergence between the variational distribution $q(\vz\mid\tilde{\vy}_s,\vc;\vphi_s)$ and the conditional prior $p(\vz\mid\vc;\vpsi_s)$.
A key difference to the standard \ac{VAE} framework~\cite{Kingma2019} is the presence of this conditional prior, which is subject to learning.
The modeled distributions are given by
\begin{equation}
    p^{(s)}(\vz\mid \vc;\vpsi_s) = \mathcal{N}(\vmu^{(s)}_\vpsi(\vc),\diag(\vsig^{(s),2}_\vpsi(\vc))) \,,
\end{equation}
\begin{equation}
    q^{(s)}(\vz\mid\tilde{\vy}_s,\vc;\vphi_s) = \mathcal{N} (\vmu^{(s)}_\vphi(\tilde{\vy}_s,\vc), \diag(\vsig^{(s),2}_\vphi(\tilde{\vy}_s,\vc)))\,,
\end{equation}
\begin{equation}
    \text{and } \; p^{(s)}(\tilde{\vy}_s\mid\vz,\vc;\vth_s) = \NC(\vmu^{(s)}_\vth(\vz,\vc),\tilde{\mc}^{(s)}_\vth(\vz,\vc))\,,
\end{equation}
where each conditional distribution is modeled as a Gaussian with parameterized mean and covariance matrix.
We model $p^{(s)}(\tilde{\vy}_s\mid\vz,\vc;\vth_s)$ as conditionally Gaussian with the second order moment $\tilde{\mc}^{(s)}_\vth(\vz,\vc)=\mc^{(s)}_\vth(\vz,\vc) + \sigma^2\I$, cf.~\cite{Baur2024}.
The \ac{cVAE} is trained by optimizing the \ac{ELBO} and by exploiting the reparameterization trick~\cite{Kingma2019}.
During training, the expectation in~\eqref{eq:elbo_h} is approximated using Monte Carlo sampling with a single draw $\vz\sim q^{(s)}(\vz\mid\tilde{\vy}_s,\vc;\vphi_s)$.

\begin{figure}
\centering
    \includegraphics[]{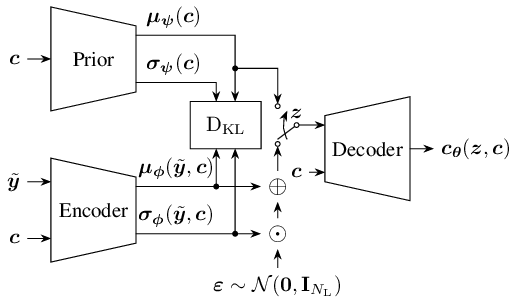}
\caption{Illustration of the \ac{cVAE}. The encoder, decoder, and prior networks are implemented using \acp{NN}. During training and inference, the latent variable $\vz$ is inferred via the decoder network (switch in the lower position) and the prior network (switch in the upper position), respectively.}
	\label{fig:cvae}
\end{figure}

When \acp{ULA} are employed at the \acp{AP}, the conditional distribution at the \ac{cVAE} output is zero-mean and has a covariance matrix with Toeplitz structure, as shown in~\cite{Boeck2024}.
Thus, we set $\vmu_\vth(\vz,\vc)=\bm{0}$.
Further, we parameterize the output covariance matrix as 
\begin{align}
   \mc^{(s)}_\vth(\vz,\vc) = \mq_{N_s}^\He\diag(\vc^{(s)}_\vth(\vz,\vc))\mq_{N_s}
\end{align}
where $\mq_{N_s}\in\mathbb{C}^{2N_s\times N_s}$ contains the first $N_s$ columns of the $2N_s\times 2N_s$ \ac{DFT} matrix, cf.~\cite{Baur2024a}.
The resulting \ac{cVAE} structure is shown in~\cref{fig:cvae}, where each block is implemented using a \ac{NN} as detailed in \cref{sec:implementation}.

The \ac{cVAE}'s output can be interpreted as a conditionally Gaussian local parameterization of $p^{(s)}(\vh_s\mid\vc)$ given by
\begin{align}
    \vh_s\mid\vz,\vc \sim p^{(s)}(\vh_s\mid\vz,\vc;\vth_s).
\end{align}
We utilize the learned prior network to approximate
\begin{align}
    p^{(s)}(\vz\mid\vc) = 
     \begin{cases}
        1 \quad\text{if } \vz = \vmu^{(s)}_\vpsi(\vc), \\
        0 \quad\text{otherwise},
    \end{cases}
\end{align}
which enables us to obtain the local conditionally Gaussian channel statistics solely from the context information as
\begin{align}
    \vh_s\mid\vc \sim \NC(\bm{0},\mc^{(s)}_\vth(\vmu^{(s)}_\vpsi(\vc),\vc)).
\end{align}
The usage of the prior network corresponds to the switch in~\cref{fig:cvae} being in the upper position, which connects the latent variable $\vz$ directly to the conditional mean of the prior network.

\subsection{Access Point Selection and Precoding}

Let us recall that the goal is to select the best \ac{AP} while maximizing its corresponding rate.
The solution to this problem can be divided into two steps. 
Firstly, we optimize the rate for each of the $S$ \acp{AP} by designing the optimal precoding strategy, i.e., calculating $\vv_s$ for all $s\in\mathcal{S}$.
Secondly, we select the \ac{AP} that yields the highest user rate $R_s$.

We do not have instantaneous \ac{CSI} available for designing the linear precoders $\vv_s$ for all \acp{AP}.
Hence, we rely on the statistics provided by the \ac{cVAE} and choose the precoders based on eigenbeamforming~\cite{Ivrlac2001}, setting $\vv_s$ to the scaled principal eigenvector of the covariance matrix $\mc^{(s)}_\vth(\vmu^{(s)}_\vpsi(\vc),\vc)$.
In principle, other choices, such as the \ac{SWMMSE} algorithm from~\cite{Razaviyayn2013}, are also possible.

The selection of the \ac{AP} that provides the optimal rate based on the designed precoders $\vv_s$ is given as
\begin{align}
    s^* = &\arg\max_s\; \log_2\left(1 + \frac{\rho_s|\vh^{\He}_s\vv_s|^2}{\sigma^2\|\vv_s\|^2}\right).\label{eq:su_ideal}
\end{align}
If the true channels $\vh_s$ were known, the rate in~\eqref{eq:su_ideal} could be directly evaluated.
Instead, we evaluate the expected rate given as
\begin{align}
    s^* = &\arg\max_s\;\mathbb{E}_{p^{(s)}(\vh_s\mid\vc)}\left[\log_2\left(1 + \frac{\rho_s|\vh^{\He}_s\vv_s|^2}{\sigma^2\|\vv_s\|^2}\right)\right].
\end{align}
To solve this optimization problem, we use the Jensen inequality to swap the logarithmic function and the expectation operator and instead evaluate an upper bound on the expected rate given as
\begin{align}
    s^* = &\arg\max_s\; \log_2\left(1 + \frac{\rho_s}{\sigma^2}\lambda_{\max}(\mc^{(s)}_\vth(\vmu^{(s)}_\vpsi(\vc),\vc))\right)\label{eq:su_upper},
\end{align}
with $\lambda_{\max}(\cdot)$ returning the maximium eigenvalue of the input matrix.
In~\eqref{eq:su_upper} we use the fact that $\vv_s$ is the eigenvector corresponding to $\lambda_{\max}$.

\subsection{Implementation Details}
\label{sec:implementation}

The simulation code is publicly available.\footnote{https://github.com/franzweisser/cvae-context-based-prediction} 
Nevertheless, we briefly outline the key aspects below.

Based on the tendency of \acp{NN} to learn low-frequency functions as observed in~\cite{Rahaman2019}, we utilize positional encoding for the context information in $\vc$.
To this end, we transform each entry $c_i$ of $\vc$ to a higher-dimensional space as
\begin{multline}
    \Bar{\vc}_i = [f_\mathrm{s}(c_i, 0),\dots,f_\mathrm{s}(c_i, K-1),\\
    f_\mathrm{c}(c_i, 0),\dots,f_\mathrm{c}(c_i, K-1)]^\T \in \mathbb{R}^{2K}
\end{multline}
with
\begin{align}
    f_\mathrm{s}(c_i, k) = \sin\left(2\pi \frac{c_i}{T_\mathrm{min}}\left(\frac{T_{\min}}{T_{\max}}\right)^{\frac{k}{K-1}}\right),\\
    f_\mathrm{c}(c_i, k) = \cos\left(2\pi \frac{c_i}{T_\mathrm{min}}\left(\frac{T_{\min}}{T_{\max}}\right)^{\frac{k}{K-1}}\right),
\end{align}
where $K$ is the number of sampled frequencies and $T_{\min}$ and $T_{\max}$ are hyperparameters.

For the encoder and decoder networks, we base our implementation on the structure proposed in~\cite{Baur2024}, where the real and imaginary parts of the \ac{CSI} input at the encoder are mapped to a higher number of convolutional channels ($N_\mathrm{C}$) using a $1\times1$ convolutional layer.
After this block, we append the encoded context information $\Bar{\vc}$ at each entry as additional convolutional channels, effectively increasing the number of convolutional channels.
Afterwards, we use again $1\times 1$ convolutional kernels to map to a representation with $N_\mathrm{C}$ convolutional channels.
The rest of the implementation follows~\cite{Baur2024}, where three blocks of a convolutional layer, a batchnormalization layer, and a ReLU activation function follow each other.
Each block increases the number of convolutional channels.
Lastly, a linear layer maps to the respective outputs in the latent space.

For the decoder input, we append the context information to the latent realization $\vz$, increasing the overall decoder's input size.
The remaining decoder network mirrors the encoder network's structure as in~\cite{Baur2024}.
The last layer is a linear layer that maps to the decoder network's outputs.

In contrast to the original work in~\cite{Sohn2015}, the prior network does not share any structure or weights with the encoder.
We implement the prior network using two linear layers, with a hidden dimension that is $8$ times the latent space dimension, and include a batch normalization layer and the ReLU activation function after the first linear layer.

\section{Simulations}

\subsection{Baseline Methods}

We compare the proposed statistical \ac{CSI} prediction technique with the following baseline methods.

In~\cite{Weisser2025c}, a \ac{CGMM} is constructed to infer channel statistics based on the user's position. 
We extend this by including the whole context information $\vc$ as input to the \ac{CGMM}.
Additionally, we compare 
to a geometric approach that simply chooses the closest, unblocked \ac{AP} and transmits solely over the \ac{LOS} path, as proposed in~\cite{Maiberger2010, Miao2021}. We denote this approach by \texttt{Geo-LoS}.
We also compare to two variants of the \ac{cVAE}.
The first variant neglects the blocking position \( \bm{b} \) and solely uses the user position \( \bm{r} \) as context information (\texttt{cVAE-pos}).
The second variant uses the latent space representation obtained from the encoder network, based on $\tilde{\vy}$ and $\vc$,
which requires additional knowledge of every \ac{CSI} for evaluation (\texttt{cVAE-csi}).

\subsection{Performance Evaluations}

We compare our proposed method, denoted as \texttt{cVAE-context}, to the baseline techniques in terms of \ac{AP} selection accuracy and the normalized rate.
For the latter, we normalize the rate obtained with a given \ac{CSI} prediction and \ac{AP} selection technique by the rate achieved with perfect \ac{CSI} knowledge.

For the training, we simulate the channel responses at each \ac{AP} for $L=1.8 \cdot 10^5$ different user and blocking positions uniformly distributed across the room using the indoor model introduced in~\cref{sec:indoor}.
For the positional encoding, we use $K=8$ with $T_{\min}=\qty{0.1}{m}$ and $T_{\max}= \qty{20}{m}$.
Further, we set $N_\mathrm{C}=16$ and $N_\mathrm{L}=16$.
During testing, we evaluate three different test scenarios $\mathcal{T}_1$, $\mathcal{T}_2$, and $\mathcal{T}_3$, each containing $10^3$ unseen tuples of user and blocking positions:
\begin{itemize}
    \item The set $\mathcal{T}_1$ contains uniformly distributed user and blocking object positions across the whole environment.
    \item The set $\mathcal{T}_2$ contains only samples where the blocking object certainly affects the choice of the optimal \ac{AP} for the case of perfect \ac{CSI}.
    \item The last set, $\mathcal{T}_3$, contains samples where the user positions are closer to \ac{AP} $2$ but behind the partition wall, as marked in~\cref{fig:sionna_indoor}.
\end{itemize}

 \begin{figure}
\centering
  \subfloat[\texttt{Perfect CSI}]{
	\centering
    \includegraphics[]{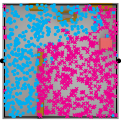}
    \includegraphics[]{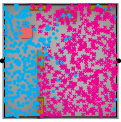}
   	}%
  \subfloat[\texttt{cVAE-context}]{
	\centering
    \includegraphics[]{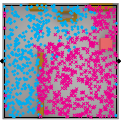}
    \includegraphics[]{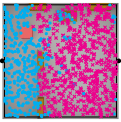}
   	}%
    \caption{AP selection performed for two possible moving object positions based on (a) perfect CSI and (b) the proposed cVAE framework. Each color (\textcolor{magenta}{red} and \textcolor{cyan}{blue}) represents user positions associated with ``AP1'' (right) and ``AP2'' (left), respectively.}
    \label{fig:ap_selection}
\end{figure}

\cref{fig:ap_selection} shows a visualization of the selected \acp{AP} for two possible blocking positions. 
The proposed method accurately captures the shadowing effect of the blocking object, selecting an alternative \ac{AP} for users in the shadowed region.
This highlights the effectiveness of the \ac{AP} selection based on the predicted statistical \ac{CSI} by the proposed \texttt{cVAE-context} method.

\begin{table}[]
\footnotesize
    \centering
    \caption{\ac{AP} selection accuracy of different \ac{CSI} prediction methods for all three test sets.}
    \scalebox{1}{
    \begin{tabular}{c|c|c|c|c}
        Method & \texttt{cVAE-context}  & \texttt{cVAE-pos} & \texttt{CGMM} & \texttt{Geo-LoS} \\\hline
        $\mathcal{T}_1$ & \textbf{0.945}  & 0.928 & 0.897 & 0.836\\\hline
        $\mathcal{T}_2$ & 0.617  & 0.246 & 0.513 & \textbf{0.866}\\\hline
        $\mathcal{T}_3$ & \textbf{0.850}  & 0.846 & 0.819 & 0.166
    \end{tabular}}
    \label{tab:su_acc}
\end{table}

Additionally, in~\cref{tab:su_acc}, the \ac{AP} selection accuracy is given for the three test cases.
The proposed \ac{CSI} prediction framework yields the most accurate \ac{AP} selection among all considered methods for $\mathcal{T}_1$ and $\mathcal{T}_3$, with $94.5\%$ and $85.0\%$ of samples correctly classified, respectively.
In both cases, the \texttt{cVAE-pos} method closely follows the proposed technique.
This is because the majority of achievable rates for the test samples are unaffected by the blocking object.
Hence, the additional input, i.e., the blocking object's position, does not significantly improve the performance.
Furthermore, the \texttt{Geo-LoS} method fails for $\mathcal{T}_3$, as it ignores the environment's geometry and therefore does not account for the partition wall and its effect on the radio propagation.
For $\mathcal{T}_2$, \texttt{Geo-LoS} performs best, as it correctly captures the effect of the moving obstacle on the \ac{LOS} path. In contrast, \texttt{cVAE-pos} shows lower accuracy since it is unaware of the moving object's position and, consequently, of its effect on the optimal \ac{AP} selection.

\begin{figure*}[t]
\centering
\includegraphics[]{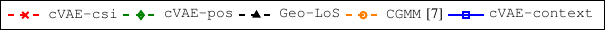}
  \subfloat[Test set $\mathcal{T}_1$]{
	\centering
    \includegraphics[]{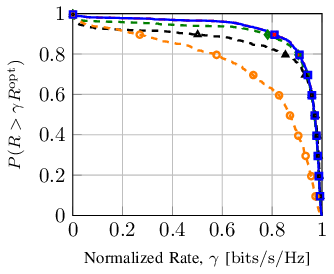}
    \label{fig:cdf_sr_su_10dB_t1}
   	}%
    \subfloat[Test set $\mathcal{T}_2$]{
	\centering
    \includegraphics[]{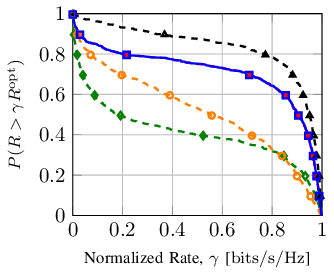}
    \label{fig:cdf_sr_su_10dB_t2}
   	}%
  \subfloat[Test set $\mathcal{T}_3$]{
	\centering
    \includegraphics[]{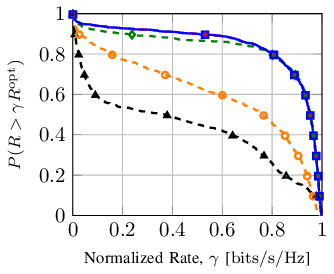}
    \label{fig:cdf_sr_su_10dB_t3}
   	}%
 \caption{Empirical cCDFs of the normalized rate achieved with different context-based \ac{CSI} prediction and \ac{AP} selection approaches at \ac{SNR} $= \qty{10}{dB}$ for the test sets (a) $\mathcal{T}_1$, (b) $\mathcal{T}_2$, and (c) $\mathcal{T}_3$.}
 \label{fig:cdf_sr_su_10dB}
\end{figure*}

We show the empirical \ac{cCDF} of the normalized rate for all three test cases in~\cref{fig:cdf_sr_su_10dB}.
The \texttt{cVAE-context} method outperforms the other baselines for $\mathcal{T}_1$ and $\mathcal{T}_3$, where only the \ac{CSI}-aware \texttt{cVAE-csi} achieves the same performance.
This observation validates the effective training of the prior network to infer the variational latent space distribution, meaning that the prior network 
very well approximates the conditional latent space distribution of the encoder network.
Furthermore, we see how the \texttt{cVAE-pos} and \texttt{Geo-LoS} methods degrade in performance for the $\mathcal{T}_2$ and $\mathcal{T}_3$ test cases, respectively.
The drop in performance of \texttt{Geo-LoS} for $\mathcal{T}_3$ is due to always selecting AP $2$, even though the partition wall blocks the connection.
Overall, we can conclude that learning the radio propagation environment and leveraging all available context information, including user and device-free object positions, yields the best \ac{CSI} predictions and \ac{AP} selection.
Lastly, we observe that the conditional Gaussian assumption for the context information itself, as used in the \ac{CGMM}, yields inferior results across all test sets.

\section{Conclusion}

This work demonstrates how context-aware \ac{CSI} prediction and \ac{AP} selection can be achieved using a \ac{cVAE}.
The proposed technique is used to predict statistical \ac{CSI} solely based on the user's position and the position of a moving obstacle. 
The results across three representative test cases demonstrate the overall benefit of learning radio propagation characteristics and exploiting all the available contextual information.
In future work, we will investigate scenarios involving multiple \acp{UE}.
Additionally, we will extend our findings to include trajectory information of all moving users and obstacles.

\balance
\bibliographystyle{IEEEtran}
  
\bibliography{IEEEabrv,mybib}

\end{document}